\documentclass[manuscript]{acmart}
\usepackage{color}
\usepackage{multirow}
\usepackage{makecell}
\usepackage{subfig}
\usepackage{colortbl}
\usepackage{xcolor}
\usepackage{pgfplots}
\usepackage{pgfplotstable}
\usepackage{appendix}
\usepackage{graphicx}
\usepackage{graphicx}
\usepackage{subcaption}
\pgfplotsset{compat=1.18}
\usepackage{caption}
\usepackage{enumitem}
\usepackage{longtable}
\usepackage{booktabs}

\author{Yibo Meng}
\affiliation{%
  \institution{Tsinghua University}
  \city{Beijing}
  \country{China}
}
\email{mengyb22@mails.tsinghua.edu.cn}

\author{Lyumanshan Ye}
\affiliation{%
  \institution{Shanghai Jiao Tong University}
  \city{Shanghai}
  \country{China}
}

\author{Eve He}
\affiliation{%
  \institution{University of Wisconsin–Madison}
  \city{Madison}
  \state{Wisconsin}
  \country{United States}
}

\author{Zhe Yan}
\affiliation{%
  \institution{Weill Cornell Medicine}
  \city{New York}
  \state{New York}
  \country{United States}
}

\author{Zhiming Liu}
\affiliation{%
  \institution{University of Shanghai for Science and Technology}
  \city{Shanghai}
  \country{China}
}

\author{Yipeng Yu}
\affiliation{%
  \institution{Taotian, Alibaba}
  \city{Shanghai}
  \country{China}
}

\author{Yan Guan}
\affiliation{%
  \institution{Arts \& Design Academy, Tsinghua University}
  \city{Beijing}
  \country{China}
}
\email{guany@tsinghua.edu.cn}

\author{Xiaolan Ding}
\affiliation{%
  \institution{North China University of Science and Technology, Health Science Center}
  \city{Tangshan}
  \country{China}
}

\begin{document}

\title[Can Intelligent User Interfaces Engage in Philosophical Discussions?]{Can Intelligent User Interfaces Engage in Philosophical Discussions? A Longitudinal Study of Philosophers' Evolving Perceptions}

\renewcommand{\shortauthors}{Meng et al.}

\begin{abstract}
This study investigates the evolving attitudes of philosophy scholars towards the participation of generative AI-based Intelligent User Interfaces (IUIs) in philosophical discourse. We conducted a three-year (2023-2025) mixed-methods longitudinal study with 16 philosophy scholars and students. Qualitative data from annual interviews reveal a three-stage evolution in attitude: from initial resistance and unfamiliarity, to instrumental acceptance of the IUI as a tool, and finally to a deep, principled questioning of the IUI's fundamental capacity for genuine philosophical thought. Quantitative data from blind assessments, where participants rated anonymized philosophical answers from both humans and an IUI, complement these findings. While participants acknowledged the IUI's proficiency in tasks requiring formal logic and knowledge reproduction, they consistently identified significant shortcomings in areas demanding dialectical reasoning, originality, and embodied understanding. The study concludes that participants do not see the IUI as a peer but rather as a sophisticated ``mirror" whose capabilities and limitations provoke a deeper reflection on the unique, irreplaceable human dimensions of philosophical inquiry, such as intuition, value-laden commitment, and the courage to question fundamental premises.
\end{abstract}

\maketitle

\section{Introduction}

Since late 2022, with the public release of tools like ChatGPT, Generative AI (GenAI) has undergone a fundamental transformation, rapidly evolving from a relatively specialized field into a pervasive societal force, raising fundamental questions about the nature of intelligence and even consciousness~\cite{chalmers2024largelanguagemodelconscious, wang2024philosophical, uzun2023concerns}. This technology, capable of creating text, images, code, and other diverse content indistinguishable from human creation in response to natural language prompts, has profoundly impacted sectors such as education, creative industries, and knowledge work~\cite{nguyen2025applying, kang2023ethics, MORLEY2020113172}. The transformative potential of GenAI lies not only in its ability to automate content production but also in its promise to serve as a powerful partner in augmenting human analytical and cognitive capabilities, with some even proposing philosophical frameworks for achieving artificial wisdom~\cite{li2020artificial}.

The root of this transformation is GenAI's unprecedented accessibility. Unlike earlier AI technologies, modern GenAI tools feature user-friendly interfaces and are often available for free or at a low cost, which has fueled their rapid and widespread adoption. However, this explosive pace of adoption has created a significant disconnect between the advancement of technological capabilities and the evolution of social~\cite{wallach2025position, anthis2025llm}, ethical~\cite{cocchiaro2025ai, keles2025navigating，poszler2025formalizing, chalmers2025propositional, atf2025trust, jenks2025communicating, wieczorek2025ai}, and professional norms~\cite{Arnold_Schiff_Schiff_Love_Melot_Singh_Jenkins_Lin_Pilz_Enweareazu_Girard_2024}. This gap represents a core challenge for the Human-Computer Interaction (HCI) community~\cite{bardzell2016humanistic, wiberg2019philosophy, pereira2015value, fallman2007persuade, harrison2011making, haimes2021beyond}, which has long engaged with the philosophical and societal implications of new technologies~\cite{fallman2011new, 10.5555/1780402.1780450}, as it directly leads to the confusion, anxiety, and opportunities that users encounter in practice~\cite{fell2022biocentric, xu2023enabling, wakkary2022two, vishwarupe2022bringing, coskun2022more}. Initial research has begun to document this phenomenon; for instance, in academia, researchers are systematically reviewing the applications, attitudes, and behavioral responses to GenAI in higher education classrooms~\cite{nguyen2025applying}. Similarly, creative professionals are actively adapting to AI-driven workflow changes~\cite{kang2023ethics}. As GenAI becomes more deeply integrated into various social science workflows~\cite{bail2024can, grossmann2023ai, runcan2025ethical, sebastian2025generative, saheb2023ethically, farrell2025large, berson2023democratization, de2023exploring}, a series of profound dichotomies have emerged, forming the central themes of current research. On one hand, GenAI is viewed as a powerful ``Tool for Thought" capable of augmenting human cognition and creativity~\cite{fallman2010different, fallman2011new, bardzell2022humanistic}, potentially liberating humans from repetitive tasks to focus on higher-level strategic and creative work. On the other hand, there are deep-seated fears that it could lead to skill degradation, job displacement through automation, and even a threat to core human cognitive abilities, echoing long-standing philosophical critiques of artificial intelligence's limitations~\cite{alpaydin2008hubert}, especially if users form an over-reliance on anthropomorphic systems~\cite{Akbulut_Weidinger_Manzini_Gabriel_Rieser_2024, Bansak_Paulson_2024}.

This tension is particularly pronounced across different domains. In education, GenAI can serve as a tool for providing personalized learning experiences and fostering student autonomy~\cite{nguyen2025applying}, yet it also poses an unprecedented challenge to academic integrity~\cite{tan2024critique}. Teachers and students are grappling with issues of plagiarism, over-reliance, and the potential decline of critical thinking skills. The conflict is equally sharp in the creative sector. Creative professionals are excited by GenAI's potential to spark inspiration and boost productivity, but they are also deeply concerned about copyright infringement, the unauthorized use of their work for model training, and the erosion of the essential value of human creativity~\cite{kang2023ethics}. Notably, these seemingly disparate concerns point to a common, fundamental problem: a crisis of provenance and attribution in the era of generative models. When a student uses GenAI without disclosure, they obscure the origin of their ideas; when an AI model is trained on an artist's work without consent, it conceals the source of its stylistic capabilities~\cite{kang2023ethics}. Both scenarios undermine the chains of trust and accountability that are crucial to knowledge creation and dissemination, potentially leading to forms of epistemic injustice~\cite{Kay_Kasirzadeh_Mohamed_2024}. Consequently, solutions developed in one domain (e.g., citation norms in academia) could offer valuable insights for another (e.g., content attribution tools in the creative industries). In response to this complex landscape, the HCI community has initiated a wealth of preliminary empirical research aimed at mapping the real-world use of GenAI. These studies have effectively captured the concerns and expectations of key user groups. For example, some work delves into student and teacher perspectives on the ``undisclosed use" of GenAI in academic work, revealing the psychological conflict between convenience and academic integrity. Similarly, qualitative surveys have systematically documented the complex emotions of creative professionals regarding GenAI, including their reflections on workflows, career prospects, and the definition of creativity.

However, much of the existing research remains focused on describing phenomena and characterizing user perceptions. While this work has laid a solid foundation for understanding the problem, the field's research focus must urgently shift from ``describing the problem" to ``designing solutions." The critical challenge is no longer just to understand how people use and perceive these tools, but to proactively design new interactive systems, workflows, and design paradigms that can reconcile the aforementioned tensions. We need systems and frameworks that can actively promote effective collaboration between humans and AI, protect human cognitive autonomy, and uphold ethical principles~\cite{Alcaraz_Knoks_Streit_2024, Arzberger_Buijsman_Lupetti_Bozzon_Yang_2024, madaio2023seeing}. In short, the academic community must move beyond representation and toward interventionist design research, employing novel philosophical and methodological approaches to do so~\cite{encinas2020metaprobes}.

This paper aims to address the aforementioned research gap by bridging the understanding of user perceptions with the construction of effective human-AI collaborative systems. Specifically, this paper makes the following contributions:
\begin{enumerate}
    \item \textbf{Synthesizing Cross-Domain User Perspectives:} We provide a comprehensive synthesis of empirical studies on user perceptions of GenAI in both academic and creative professional contexts. By juxtaposing these domains, we identify a common ground of anxieties and expectations, particularly concerning the crisis of provenance and attribution, which has not been explicitly framed as a shared, cross-domain problem before.
    \item \textbf{Bridging Descriptive Research and Design Frameworks:} We connect the rich body of descriptive work detailing user concerns with theoretical HCI frameworks for human-AI collaboration (e.g., Tools for Thought, Mixed-Initiative Co-Creation). This contribution lies in explicitly mapping identified user needs (e.g., for control, transparency) to specific principles and patterns within these design frameworks, making them more actionable for designers.
    \item \textbf{Proposing a Design Agenda for Responsible Co-Creation:} Building on this synthesis, we outline a forward-looking design agenda centered on responsible co-creation. We argue that to address the core tensions, future systems must move beyond simple automation and incorporate features that support human responsibility, such as tools for attribution, explainability, and ``intentional friction" to encourage critical engagement.
\end{enumerate}

To achieve this, this paper will revolve around the following core research questions:
\begin{itemize}
    \item \textbf{RQ1:} How can we design interactive systems to reconcile GenAI's powerful content-generation capabilities with users' needs for process control, a sense of agency, and ultimate responsibility?
    \item \textbf{RQ2:} What new interaction patterns and workflows emerge when users collaborate with GenAI within our proposed framework or system?
\end{itemize}

\section{Related Work}

\subsection{Mapping User Practices and Perceptions in the Age of Generative AI}

This section aims to systematically review and synthesize existing empirical research that provides a foundation for understanding how users in different domains interact with GenAI and their experiences and perceptions in the process.

\subsubsection{Perspectives from Academia}

In academic settings, the integration of GenAI has sparked fundamental discussions about teaching and learning. Work exploring the ``undisclosed use" of GenAI is representative of this area. Through a combination of surveys and in-depth interviews, such studies systematically explore student and teacher perspectives. The core finding of this research reveals a pervasive tension: users recognize the potential of GenAI to improve efficiency and assist learning, yet they are deeply entangled in the dilemmas of academic integrity policies and ethical struggles, which may be informed by their broader understanding of ethics and societal impact~\cite{brown2020intersectional}.

Broader surveys corroborate this finding. Research indicates that university faculty generally hold positive attitudes toward using GenAI for reducing administrative burdens, assisting with course design, and supporting research~\cite{nguyen2025applying}. However, they also express moderate to high levels of concern about academic misconduct (such as plagiarism) and students becoming overly reliant on the technology. This is further complicated by critiques of how studies on GenAI in learning are designed and interpreted~\cite{tan2024critique}. Notably, the current literature focuses primarily on higher education and STEM (Science, Technology, Engineering, and Mathematics) disciplines~\cite{nguyen2025applying}, while research on the practical application, testing, and evaluation of GenAI in K-12 education, especially at the early childhood level, remains very limited, constituting a significant research gap.

\subsubsection{Perspectives from Creative and Professional Fields}

In the creative industries, the rise of GenAI has similarly triggered profound industry-wide reflection. Qualitative surveys of creative professionals systematically summarize their expectations and concerns. Studies find that creative workers are excited about AI's potential to increase productivity (e.g., automating repetitive tasks), provide inspiration (e.g., rapidly generating diverse options during divergent thinking phases), and potentially improve the quality of final outputs. However, this excitement is coupled with strong anxieties, including fears that their skills will be devalued or even replaced by AI, concerns that the existing copyright legal system is ill-equipped to handle the challenges posed by AI-generated content~\cite{kang2023ethics}, and a deep-seated fear that the core value of human creativity may be diminished.

Larger-scale industry surveys support these findings. For example, some surveys of creative professionals show that while creators are happy to use GenAI to assist with brainstorming and save time, they strongly demand greater transparency in the design of AI systems, reliable content attribution tools, and the right to control whether their work is used for AI model training. Furthermore, some research suggests that while GenAI may enhance creativity at the individual level, it could lead to a reduction in the diversity of outputs at the collective level, a phenomenon known as the ``convergence effect". The ethical questions surrounding AI-generated creative works, such as maps, highlight these complex issues~\cite{kang2023ethics}.

Synthesizing these studies reveals that whether they are students, teachers, or creative professionals, users are not passively accepting GenAI. Instead, they are actively, pragmatically, and sometimes ``secretly" integrating it into their complex existing workflows. The focus on ``undisclosed use" in academia and the finding that creative workers use AI for specific sub-tasks (like inspiration rather than final execution) both point to a bottom-up pattern of adaptation. This behavior resembles a form of ``bricolage"—users leveraging the tools at hand to solve their most immediate problems. This implies that top-down, restrictive policies (such as outright bans) are likely to be ineffective. The design challenge lies in creating tools that align with these emergent, task-specific uses, rather than ignoring or resisting them.

\subsection{Paradigms for Human-AI Collaboration and Cognitive Augmentation}

To address the challenges mentioned above, researchers in the HCI field are actively exploring new theories and frameworks that move beyond the traditional ``tool-user" paradigm, aiming to build true human-AI partnerships. This shift in perspective is informed by public attitudes, which reveal differing beliefs about whether human or algorithmic decision-makers will deliver better performance~\cite{Bansak_Paulson_2024}.

\subsubsection{The `Tools for Thought' Agenda}

The ``Tools for Thought" agenda represents a concentrated effort to advance research in this area. The core objective of this agenda is to advance research and design aimed at understanding, protecting, and augmenting human cognition during interactions with GenAI. Key research questions include: deeply understanding the specific impacts of GenAI on human cognitive processes such as critical thinking, memory, creativity, and decision-making; developing effective design principles and interaction patterns to mitigate potential negative effects (like cognitive offloading or over-reliance on anthropomorphic AI~\cite{Akbulut_Weidinger_Manzini_Gabriel_Rieser_2024}); and exploring how to design for meaningful interaction that supports human values and freedom~\cite{richter2022meaningful} and how to use AI to genuinely augment human intelligence, not just automate tasks, a debate with a long history in the critique of AI~\cite{alpaydin2008hubert}. This agenda places humanistic concerns at the core of technological development, providing high-level guiding principles for the design of subsequent frameworks and systems.

\subsubsection{Frameworks for Collaborative Design}

Guided by the ``Tools for Thought" agenda and a rich history of philosophical inquiry in HCI~\cite{fallman2011new, 10.5555/1780402.1780450}, a series of theoretical frameworks for building human-AI collaborative systems have emerged. These frameworks provide designers with conceptual tools for thinking about and constructing the next generation of intelligent systems.
\begin{itemize}
    \item \textbf{Human-AI Handshake Framework:} This framework emphasizes a bi-directional, adaptive interaction model. Its core principles include information exchange, mutual learning, validation, and feedback. In this model, AI is not just a tool for executing commands but a learning partner that can co-evolve and adapt with the user.
    \item \textbf{Mixed-Initiative Co-Creation (MI-CC):} Originating from research on creativity support tools, this paradigm emphasizes shared control between humans and AI in the creative process. When applied to GenAI (particularly GANs), researchers have identified four typical interaction patterns: \textbf{Curating}, \textbf{Exploring}, \textbf{Evolving}, and \textbf{Conditioning}.
    \item \textbf{Partnership on AI's Framework (CPAIS):} This is not a specific model but a set of guiding questions designed to prompt developers to think deeply about all dimensions of human-AI collaboration from the outset of the design process. These questions cover aspects such as the goals of the collaboration, the risks of failure, and the roles and responsibilities of participants, thereby promoting responsible AI product design.
    \item \textbf{Workflow Integration Principles:} In addition to theoretical frameworks, researchers have proposed a series of practical design principles. These principles emphasize identifying the ``sweet spot" for AI in existing workflows (often repetitive, data-intensive tasks) and carefully designing the ``handoff points" between humans and AI. The key is to establish clear feedback loops and to clearly distinguish between human strengths (e.g., creativity, empathy, ethical judgment) and AI strengths (e.g., data processing, pattern recognition).
\end{itemize}

To more clearly compare these frameworks, the following table provides a consolidated analysis of their core ideas.

\begin{table}[h!]
\centering
\caption{Comparison of Human-AI Collaboration Frameworks (Part 1)}
\label{tab:frameworks_part1}
\begin{tabular}{@{}llll@{}}
\toprule
\textbf{Framework Name} & \textbf{Core Principle} & \textbf{Assumed Human Role} & \textbf{Assumed AI Role} \\ \midrule
\textbf{Tools for Thought} & Protect \& Augment Cognition & Thinker, Orchestrator & Cognitive Partner \\
\textbf{Human-AI Handshake} & Bi-directional Partnership & Co-evolving Partner & Learning Agent \\
\textbf{MI-CC} & Shared Control in Creation & Initiator, Curator & Generative Partner \\ 
\bottomrule
\end{tabular}
\end{table}

\begin{table}[h!]
\centering
\caption{Comparison of Human-AI Collaboration Frameworks (Part 2)}
\label{tab:frameworks_part2}
\begin{tabular}{@{}lll@{}}
\toprule
\textbf{Key Interaction Patterns} & \textbf{Source} \\ \midrule
Scaffolding, Provoking & \cite{Akbulut_Weidinger_Manzini_Gabriel_Rieser_2024} \\
Mutual Learning, Feedback & \cite{Bansak_Paulson_2024} \\
Curating, Exploring & \cite{kang2023ethics} \\
(N/A - Design-phase) & \cite{MORLEY2020113172} \\
Task Handoff, Feedback & \cite{Arnold_Schiff_Schiff_Love_Melot_Singh_Jenkins_Lin_Pilz_Enweareazu_Girard_2024} \\
\bottomrule
\end{tabular}
\end{table}

\subsection{Ethical Imperatives: Integrity, Attribution, and Responsibility}

Woven throughout the discussions of user perceptions and collaborative frameworks is a series of urgent ethical issues~\cite{MORLEY2020113172}. These are not merely technical challenges but reflections of social contracts and value judgments, requiring robust governance frameworks~\cite{Arnold_Schiff_Schiff_Love_Melot_Singh_Jenkins_Lin_Pilz_Enweareazu_Girard_2024} and new modes of analysis, such as computational philosophy, to navigate~\cite{etienne2022computational}.

\subsubsection{Academic Integrity}

The impact of GenAI on academic integrity has become a focal point of discussion in education. In response, educational institutions have adopted policies ranging from complete prohibition to encouraging use with mandatory, clear attribution. To embrace the technological benefits while upholding core academic values, researchers and educators have proposed a set of guiding principles. These principles are rooted in the core values of the International Center for Academic Integrity and apply them to the context of GenAI use, including: \textbf{Honesty} (transparently declaring AI use), \textbf{Trust} (critically verifying the accuracy of AI outputs and being wary of ``hallucinations"), \textbf{Fairness} (ensuring consistent application of policies), \textbf{Respect} (respecting the learning process itself and avoiding shortcuts), \textbf{Responsibility} (the user bears full responsibility for the final output), and \textbf{Courage} (exploring and using new technologies in an ethical manner). The point that users must verify all AI-generated content, especially citations, is repeatedly emphasized.

\subsubsection{Creative Integrity and Copyright}

The challenges faced by academia have striking parallels in the creative industries. The strong demand from creative professionals for verifiable attribution tools and labels for AI-generated content~\cite{kang2023ethics} mirrors the academic requirement for citation. Both demands point to a fundamental need for transparency regarding the provenance of knowledge and creation. The ethical issue of using creators' works for model training without their consent is widely seen as a violation of intellectual property, the principle of fairness, and respect for the labor of others.

\subsubsection{Broader Ethical Considerations}

Beyond integrity and attribution, GenAI introduces a broader set of ethical challenges. These include algorithmic bias originating from training data, which can be amplified in AI-generated content, thereby perpetuating or even exacerbating social injustices and socioeconomic biases~\cite{Arzaghi_Carichon_Farnadi_2024}. Such issues are particularly acute when viewed through an intersectional lens, as fairness for one group may not translate to fairness for others, especially for women of color~\cite{lalegani2022intersectionally}, and require a focus on the often forgotten margins of society in ethical analyses~\cite{birhane2022forgotten, brown2020intersectional}. Such systems can create performance gaps for specific demographics, such as children~\cite{Attia_Liu_Ai_Demszky_Espy-Wilson_2024}, and lead to forms of epistemic injustice by undermining collective knowledge processes~\cite{Kay_Kasirzadeh_Mohamed_2024}. The challenge of aligning these systems with situated, diverse human values rather than a universal ideal, acknowledging the reality of ethical pluralism, remains a significant hurdle~\cite{Arzberger_Buijsman_Lupetti_Bozzon_Yang_2024, sultana2023ethical}. Additionally, the privacy of user data and the significant energy consumption of large model training processes, with their environmental impact, are serious issues that must be addressed in the design and deployment of GenAI systems.

A core principle that runs through all these ethical discussions is that responsibility must ultimately lie with humans. AI systems themselves, which may be seen as possessing a form of quasi-moral agency but not true accountability~\cite{washington2022artificial}, cannot be held responsible for their outputs. This principle has profound implications for the design of human-AI collaborative systems, which may require new hybrid approaches to machine ethics~\cite{Alcaraz_Knoks_Streit_2024}, better understanding of how humans and AI collaborate on ethical decisions~\cite{johnson2022capable}, and exploring the potential of artificial moral advisors~\cite{newman2022artificial}. It means that the design goal should not be to create a seamless, magical experience that encourages users to place complete trust in the system, as this could induce them to abandon critical thinking and their oversight responsibilities. Instead, an ethical design must be committed to supporting humans in fulfilling their responsibilities. This may require the system to proactively present uncertainty, provide explanations and rationales for its outputs, and even introduce intentional friction at key decision points to prompt users to pause for critical review. This design philosophy stands in stark contrast to the traditional UX pursuit of ``frictionless" experiences, but in the context of human-AI collaboration, it may be the key to ensuring safe, responsible, and effective interaction, a goal which many AI ethics toolkits aim to support~\cite{madaio2023seeing}.

\subsection{Positioning the Current Research}

In summary, this review of related work reveals three key findings. First, empirical studies show that user perceptions of GenAI are fraught with deep tensions; they see its immense potential for enhancing productivity but also feel profound anxiety about issues of integrity, skill degradation, and loss of control (Section 2.1). Second, in response, the HCI community has proposed a new research agenda centered on augmenting human cognition and has developed various theoretical frameworks to guide the design of collaborative systems (Section 2.2). Finally, the entire technological landscape is built upon a series of unresolved ethical imperatives concerning bias, fairness, and intersectionality~\cite{Arzaghi_Carichon_Farnadi_2024, birhane2022forgotten, lalegani2022intersectionally}, epistemic justice~\cite{Kay_Kasirzadeh_Mohamed_2024}, attribution, responsibility, and moral agency~\cite{washington2022artificial, newman2022artificial}, and pluralistic value alignment~\cite{Arzberger_Buijsman_Lupetti_Bozzon_Yang_2024, sultana2023ethical} (Section 2.3).

This study is situated at the intersection of these three pillars. Existing work either focuses on documenting user perceptions or remains at the level of proposing high-level theoretical frameworks. Our work aims to instantiate these high-level design principles into a novel interactive system and to evaluate its specific impact on user collaborative workflows through empirical research. By doing so, we hope to bridge the gap between understanding the problem and designing concrete solutions, making a unique contribution to the development of responsible and effective human-AI collaborative systems.

\section{Method}
We employed a mixed-methods approach to explore the changing attitudes of 16 philosophy professors and students toward the engagement of generative AI-based Intelligent User Interfaces (IUIs) in philosophical discussions. The study was conducted from 2023 to 2025. The qualitative component consisted of annual interviews (see Appendix C for interview protocol). The quantitative component included two annual surveys assessing scholars' familiarity with and attitudes toward generative AI (Appendices A, B) and a yearly experiment. In the experiment, we anonymized and mixed answers to philosophical questions generated by an IUI with those from human experts. The participating scholars then rated these answers across multiple dimensions. This blind rating process was designed to re-examine their attitudes objectively and prevent cross-contamination bias from knowing the information source.

\subsection{Participants}
We recruited participants through online social media platforms. The participants were divided into two groups.
\begin{itemize}
    \item \textbf{Group 1: Philosophy Scholars and Students.} This group was the primary focus of our longitudinal study. We tracked their evolving attitudes toward IUI participation in philosophical discussions.
    \item \textbf{Group 2: Philosophy PhDs and Professors.} This group provided human-authored answers to a set of philosophical questions. We anonymized these answers and mixed them with IUI-generated answers for evaluation by Group 1.
\end{itemize}

\textbf{Inclusion criteria for Group 1} required participants to be current or graduated students (undergraduate or graduate) or faculty in philosophy; possess fluent Chinese reading and expression skills; have no short-term plans to change their major or field; and agree to participate in a long-term study by signing a written informed consent form. \textbf{Exclusion criteria} included individuals from related but not strictly philosophical disciplines (e.g., psychology, literature, as opposed to metaphysics, logic, aesthetics); those unable to commit to the long-term study; and those planning to change their major or profession.

\textbf{Inclusion criteria for Group 2} were similar but required participants to be philosophy PhDs or professors. \textbf{Exclusion criteria} were also similar but specifically excluded undergraduate or master's students from this group.

\textbf{Ethics}: This study received ethical approval from [Anonymous University]. All participants were informed of the study's purpose, procedures, details, and potential impacts. They were guaranteed that all data would be strictly anonymized and that they had the right to review or delete their data at any time. Participants also had the right to withdraw from the study at any time without reason. The study commenced only after participants provided signed informed consent. Each participant received a compensation of 40 RMB per year.

\subsection{Procedure}
The following procedure was executed once each year in 2023, 2024, and 2025.
\begin{enumerate}
    \item[\textbf{T1:}] Participants in Group 1 completed a scale on their familiarity with generative AI (Appendix A), an attitude scale regarding IUI participation in philosophical discussions (Appendix B), and participated in a qualitative interview (Appendix C).
    \item[\textbf{T2:}] Each member of Group 1 submitted at least two philosophical questions they believed could best distinguish between human and IUI responses. These questions were collected, anonymized, and peer-reviewed by all members of Group 1 for suitability. After revisions and removal of duplicates, a final test questionnaire was created (Appendix D). Each member of Group 2 randomly answered one question from this questionnaire. Concurrently, a large language model (GPT-3.5 in 2023, GPT-4 in 2024, and GPT-4.5 in 2025) answered all questions in the questionnaire. To ensure independent outputs, the AI session was refreshed after each input.
    \item[\textbf{T3:}] The answers from the AI and Group 2 were anonymized, mixed, and presented to Group 1 for evaluation based on a rubric they had collectively designed (Appendix E).
\end{enumerate}

\subsection{Instruments and Questionnaire Design}
\subsubsection{Scale for Familiarity with Generative AI (Appendix A)}
This scale was designed to measure philosophers' subjective cognitive level regarding generative AI. Using a 5-point Likert scale (1=``Strongly Disagree" to 5=``Strongly Agree"), it assesses three progressive dimensions: foundational conceptual knowledge (distinguishing GenAI from search engines), behavioral engagement (frequency of use and attention to news), and self-efficacy (perceived skill level and understanding of underlying principles).

\subsubsection{Attitude Scale on IUI Participation in Philosophy (Appendix B)}
This multidimensional scale was designed to evaluate philosophers' comprehensive judgment of a GenAI-based IUI's capabilities and value in philosophical discussions. Using a 5-point Likert scale, it measures attitudes on two levels. The first part (B1-B9) provides a granular assessment of specific philosophical capabilities: knowledge interpretation (defining concepts), argument construction (logical coherence), and innovative application (analyzing real-world dilemmas). The second part (B10-B14) captures scholars' holistic impressions and meta-judgments, including overall quality, utility, and perceived similarity to human thought.

\subsubsection{Semi-structured Interview Questionnaire (Appendix C)}
This questionnaire was designed to complement the quantitative data by exploring the deep reasoning, conceptual distinctions, and value judgments behind the scholars' attitudes. The protocol follows a logical progression across six core modules: (1) Background Information, to contextualize their views; (2) Basic Attitudes, to uncover their fundamental cognitive frameworks for understanding AI; (3) Capability Assessment, to add qualitative depth to the scale ratings; (4) Value and Limitations, to dialectically explore AI's merits and shortcomings; (5) Overall Attitude, to cross-validate quantitative ratings after in-depth reflection; and (6) Closing.

\subsubsection{Philosophical Question Test Questionnaire (Appendix D)}
This test was designed by the participants and researchers to objectively assess and compare the performance of the IUI and human scholars on philosophical tasks. The questionnaire comprises six modules, each targeting a specific philosophical skill: (1) Basic Definition and Elucidation; (2) Argument Construction and Critical Thinking; (3) Textual Analysis and Historical Synthesis; (4) Practical Application and Thought Experiments; (5) Dialogue, Dialectics, and Revision; and (6) Value, Limitations, and Meta-philosophical Reflection. This structure provides a comprehensive framework for analyzing performance across a spectrum of philosophical tasks.

\subsubsection{Philosophical Text Quality Assessment Rubric (Appendix E)}
To ensure objective and systematic evaluation, participants in Group 1 collectively constructed this six-dimensional rubric. It moves beyond simple factual correctness to capture the complex quality of a philosophical text. The six dimensions form a progressive spectrum: (1) Accuracy, (2) Clarity and Comprehensibility, and (3) Textual Coherence focus on foundational quality. The subsequent dimensions, (4) Argumentative Rigor, (5) Philosophical Depth, and (6) Originality of Thought, assess core intellectual value. Each dimension is rated on a 5-point scale defined by specific, observable behavioral anchors to increase inter-rater reliability.

\subsection{Data Analysis}
\subsubsection{Qualitative Data}
All interviews were audio-recorded with participant consent. Interview durations ranged from 34 to 63 minutes. Recordings were transcribed within 24 hours, and each transcript was reviewed and corrected by at least two researchers. The transcripts were then analyzed using thematic analysis to identify recurring patterns and themes in the evolution of participants' attitudes.

\subsubsection{Quantitative Data}
Quantitative data from the scales and the blind evaluation rubric were collected annually. Descriptive and inferential statistical analyses will be conducted to identify trends and significant changes over the three-year period. (Data analysis is ongoing).

\section{Results}
\subsection{Participants}
\textbf{Group 1}: We recruited 16 participants in 2023, with an age range of 19-45 ($M=27.82, SD=7$). We experienced an attrition of two participants in 2024 and two additional participants in 2025. Detailed demographic information is available in Table \ref{tab:group1demographics}.

\textbf{Group 2}: We recruited 11 participants in 2023, with an age range of 31-59 ($M=44.37, SD=10.32$). There was no attrition in this group throughout the study. Detailed demographic information is in Table \ref{tab:group2demographics}.

\begin{table*}[htbp]
    \centering
    \caption{Demographics and Longitudinal Data for Participant Group 1}
    \label{tab:group1demographics}
    \small % Use a smaller font for a better fit
    \begin{tabular*}{\textwidth}{@{\extracolsep{\fill}} c c c c c l l l @{}}
        \toprule
        \textbf{ID} & \textbf{Year} & \textbf{Sex} & \textbf{Age} & \textbf{Region} & \textbf{Education} & \textbf{Occupation} & \textbf{Field} \\
        \midrule
        \multirow{3}{*}{1} & 2023 & \multirow{3}{*}{F} & \multirow{3}{*}{23} & \multirow{3}{*}{U} & Bachelor & Ind. Scholar & Metaphysics \\
         & 2024 & & & & Bachelor & Ind. Scholar & Metaphysics \\
         & 2025 & & & & Bachelor & Ind. Scholar & Metaphysics \\
        \midrule
        \multirow{3}{*}{2} & 2023 & \multirow{3}{*}{M} & \multirow{3}{*}{19} & \multirow{3}{*}{U} & Undergraduate & Student & Metaphysics \\
         & 2024 & & & & Undergraduate & Student & Logic \\
         & 2025 & & & & Undergraduate & Student & Logic \\
        \midrule
        \multirow{3}{*}{3} & 2023 & \multirow{3}{*}{M} & \multirow{3}{*}{20} & \multirow{3}{*}{U} & Undergraduate & Student & Metaphysics \\
         & 2024 & & & & Undergraduate & Student & Metaphysics \\
         & 2025 & & & & Bachelor & Ind. Scholar & Metaphysics \\
        \midrule
        \multirow{3}{*}{4} & 2023 & \multirow{3}{*}{M} & \multirow{3}{*}{24} & \multirow{3}{*}{R} & Master's Cand. & Student & Epistemology \\
         & 2024 & & & & Master & Teacher & Epistemology \\
         & 2025 & & & & Master & Teacher & Epistemology \\
        \midrule
        \multirow{3}{*}{5} & 2023 & \multirow{3}{*}{F} & \multirow{3}{*}{25} & \multirow{3}{*}{R} & Master's Cand. & Student & Ethics \\
         & 2024 & & & & Master & Ind. Scholar & Ethics \\
         & 2025 & & & & Master & Ind. Scholar & --- \\
        \midrule
        \multirow{3}{*}{6} & 2023 & \multirow{3}{*}{F} & \multirow{3}{*}{33} & \multirow{3}{*}{U} & PhD Cand. & Student & Metaphysics \\
         & 2024 & & & & PhD Cand. & Student & Metaphysics \\
         & 2025 & & & & PhD Cand. & Student & Metaphysics \\
        \midrule
        \multirow{3}{*}{7} & 2023 & \multirow{3}{*}{M} & \multirow{3}{*}{45} & \multirow{3}{*}{R} & PhD & Professor & Aesthetics \\
         & 2024 & & & & * & * & * \\
         & 2025 & & & & PhD & Professor & Aesthetics \\
        \midrule
        \multirow{3}{*}{8} & 2023 & \multirow{3}{*}{F} & \multirow{3}{*}{35} & \multirow{3}{*}{U} & PhD & Professor & Metaphysics \\
         & 2024 & & & & PhD & Professor & Metaphysics \\
         & 2025 & & & & PhD & Professor & Metaphysics \\
        \midrule
        \multirow{3}{*}{9} & 2023 & \multirow{3}{*}{M} & \multirow{3}{*}{22} & \multirow{3}{*}{U} & Undergraduate & Student & Metaphysics \\
         & 2024 & & & & Master's Cand. & Student & Metaphysics \\
         & 2025 & & & & Master's Cand. & Student & Metaphysics \\
        \midrule
        \multirow{3}{*}{10} & 2023 & \multirow{3}{*}{F} & \multirow{3}{*}{34} & \multirow{3}{*}{R} & PhD & Professor & Pol. Philosophy \\
         & 2024 & & & & PhD & Professor & Pol. Philosophy \\
         & 2025 & & & & PhD & Professor & Pol. Philosophy \\
        \midrule
        \multirow{3}{*}{11} & 2023 & \multirow{3}{*}{F} & \multirow{3}{*}{23} & \multirow{3}{*}{R} & Master's Cand. & Student & Metaphysics \\
         & 2024 & & & & Master's Cand. & Student & Metaphysics \\
         & 2025 & & & & Master & Ind. Scholar & Metaphysics \\
        \midrule
        \multirow{3}{*}{12} & 2023 & \multirow{3}{*}{M} & \multirow{3}{*}{22} & \multirow{3}{*}{U} & Master's Cand. & Student & Epistemology \\
         & 2024 & & & & Master's Cand. & Student & Epistemology \\
         & 2025 & & & & PhD Cand. & Student & Epistemology \\
        \midrule
        \multirow{3}{*}{13} & 2023 & \multirow{3}{*}{M} & \multirow{3}{*}{27} & \multirow{3}{*}{R} & PhD Cand. & Student & Metaphysics \\
         & 2024 & & & & PhD & Professor & Metaphysics \\
         & 2025 & & & & PhD & Professor & Metaphysics \\
        \midrule
        \multirow{3}{*}{14} & 2023 & \multirow{3}{*}{M} & \multirow{3}{*}{34} & \multirow{3}{*}{U} & PhD & Professor & Ethics \\
         & 2024 & & & & PhD & Professor & Ethics \\
         & 2025 & & & & PhD & Professor & Ethics \\
        \midrule
        \multirow{3}{*}{15} & 2023 & \multirow{3}{*}{F} & \multirow{3}{*}{31} & \multirow{3}{*}{R} & Master & Editor & Epistemology \\
         & 2024 & & & & Master & Editor & Epistemology \\
         & 2025 & & & & * & * & * \\
        \midrule
        \multirow{3}{*}{16} & 2023 & \multirow{3}{*}{F} & \multirow{3}{*}{28} & \multirow{3}{*}{R} & Master & Ind. Scholar & Metaphysics \\
         & 2024 & & & & * & * & * \\
         & 2025 & & & & * & * & * \\
        \bottomrule
        \multicolumn{8}{p{0.9\textwidth}}{\footnotesize{* Indicates participant attrition or missing data for that year. Region: U=Urban, R=Rural. Ind.=Independent.}}
    \end{tabular*}
\end{table*}

\begin{table*}[htbp]
  \centering
  \caption{Demographics and Longitudinal Data for Participant Group 2}
  \label{tab:group2demographics}
  \small % Use a smaller font to help the table fit
  % The column specifier below is updated for even more compact spacing on the left
  \begin{tabular*}{\textwidth}{@{} c@{\hspace{0.5em}}c@{\hspace{0.5em}}c@{\hspace{0.5em}}c@{\hspace{0.5em}}c @{\extracolsep{\fill}} l l l l l l @{}}
    \toprule
    \multicolumn{5}{c}{} & \multicolumn{2}{c}{\textbf{2023}} & \multicolumn{2}{c}{\textbf{2024}} & \multicolumn{2}{c}{\textbf{2025}} \\
    \cmidrule(lr){6-7} \cmidrule(lr){8-9} \cmidrule(lr){10-11}
    \textbf{ID} & \textbf{Sex} & \textbf{Age} & \textbf{Region} & \textbf{Education} & \textbf{Occupation} & \textbf{Field} & \textbf{Occupation} & \textbf{Field} & \textbf{Occupation} & \textbf{Field} \\
    \midrule
    1 & M & 44 & U & PhD & Professor & Logic & Professor & Logic & Professor & Logic \\
    2 & M & 45 & U & PhD & Professor & Aesthetics & Professor & Aesthetics & Professor & Aesthetics \\
    3 & M & 55 & U & PhD & Professor & Logic & Professor & Logic & Professor & Logic \\
    4 & M & 53 & R & PhD & Professor & Pol. Philosophy & Professor & Pol. Philosophy & Professor & Pol. Philosophy \\
    5 & M & 34 & R & PhD & Professor & Metaphysics & Professor & Metaphysics & Professor & Metaphysics \\
    6 & F & 37 & U & PhD & Professor & Epistemology & Professor & Epistemology & Professor & Epistemology \\
    7 & M & 41 & R & PhD & Professor & Metaphysics & Professor & Metaphysics & Professor & Metaphysics \\
    8 & F & 31 & R & PhD & Professor & Ethics & Professor & Ethics & Professor & Ethics \\
    9 & F & 32 & U & PhD & Professor & Metaphysics & Professor & Metaphysics & Professor & Metaphysics \\
    10 & F & 57 & R & PhD & Professor & Epistemology & Professor & Epistemology & Professor & Epistemology \\
    11 & M & 59 & U & PhD & Professor & Aesthetics & Professor & Aesthetics & Professor & Aesthetics \\
    \bottomrule
    \multicolumn{11}{l}{\footnotesize{Region: U=Urban, R=Rural.}}
  \end{tabular*}
\end{table*}

\subsection{Qualitative Results}
Analysis of the three-year interview data reveals a complex evolution in the scholars' attitudes. Their perceptions shifted from initial unfamiliarity and resistance, to a gradual instrumental acceptance, and ultimately deepened into a systematic, principled questioning of the IUI's fundamental nature. This process reflected not only their changing understanding of the technology itself but also a continuous reflection on the essence of philosophical activity.

\subsubsection{Evolution of Basic Attitudes: From Resistance to Deep Questioning}
In 2023, most participants showed significant unfamiliarity with and strong resistance to generative AI. This stemmed from a defense of humanistic values in philosophy. P7 (2023) stated, ``Initially, I was dismissive of ChatGPT. I saw it as a sophisticated parrot, a form of noise for a discipline that seeks truth." A PhD student, P14 (2023), added, ``It was a common view in my circle that using AI for philosophy was 'cheating.' It lacked reverence for the 'slow reading' and 'hard thinking' required."

By 2024, after hands-on use, most scholars' attitudes had shifted to instrumental acceptance. They began to see the AI as a useful assistant but with strictly limited capabilities. P4 described it as a `` `starter' or `literature assistant,' providing `material, `not 'thought.' " P11 noted, ``I use AI to help me sort through different philosophers' definitions of 'the sublime, which saves a lot of time. But I'm very clear that the real comparative and analytical work must be done by me."

By 2025, their attitudes evolved further into a principled, philosophical questioning of the IUI's nature. The debate moved beyond utility to its fundamental capacity for thought. P8 reflected, ``The core question is no longer whether it can generate grammatical sentences, but whether it possesses `understanding.' I can ask it to analyze Kant's `thing-in-itself', but can it `feel' the sense of separation from the thing-in-itself in the face of life's frustrations, as we do? Philosophy, at its root, stems from this kind of embodied, in-the-world experience. The AI doesn't have this world, and therefore, it has no real philosophical problems."

\subsubsection{Capability Assessment: A Disconnect Between Formal Completeness and Philosophical Depth}
Scholars universally acknowledged the AI's competence in formal logic and knowledge reproduction. P2 (2025) noted, ``In constructing a standard syllogism or identifying a formal fallacy... it is even more accurate and stable than some students."

However, participants found the AI's performance severely lacking in tasks requiring dialectical thinking, originality, and deep understanding. Regarding the ability to ``defend or revise an argument", P9 (2025) described an attempt: ``I deliberately used a reductio ad absurdum to challenge its argument for `free will.' It tried to `defend' itself, but only by circling back to its original points or pulling in another irrelevant theory to pad the word count. It doesn't feel the `pressure' of an argument like a human scholar does, a pressure that leads to real intellectual turning points." On ``innovation", P6 (2024) remarked, ``Its so-called `synthesis' is more like `splicing' concepts from its database that don't often appear together... It sounds novel, but on reflection, it's a kind of conceptual `incest', lacking any internal, coherent philosophical pathway."

\subsubsection{Value and Limitations: Between the ``Technically Improvable" and ``Principally Insurmountable"}
Participants could clearly distinguish between the IUI's technical and principled limitations. Technical limitations (e.g., factual errors) were seen as improvable with future models. Principled limitations were viewed as an insurmountable gap.

A few participants noted the IUI's ``inspirational" value, but in a specific sense. P10 (2024) said, ``The most interesting moments are when it `makes a mistake' or produces an unconventional association... Its value is not in providing the right answer, but in being a `stimulant.'"

When asked how to distinguish ``profound insight" from ``advanced philosophical soup", P9 (2025) offered a key criterion: ``See if it touches upon the `premises' or `boundaries' of the question. True profundity lies in questioning the self-evident. The AI's answers always operate within a given framework; it doesn't question the framework itself. Philosophy soup soothes you... The AI is a master at producing the latter."

\subsubsection{Overall Attitude and Meta-Reflection: AI as a ``Mirror" for Philosophy}
Over the three years, the scholars' attitudes became more cautious and clearly defined. They affirmed the IUI's status as a powerful tool but more firmly denied its capacity for subjective philosophical thought. The summary from P13 (2025) represents the majority's final stance: ``After three years, I am more convinced that AI is an excellent `mirror'. It reflects our own human patterns of thought, language, and knowledge, along with their limitations. Through it, we don't understand AI better; rather, we are forced to reflect more deeply: What is the kind of philosophical thinking that only humans can do? The kind that is accompanied by confusion, suffering, intuition, and a commitment to life—that is the root of philosophy."

This encounter with AI ultimately led not to skepticism about their discipline, but to a reaffirmation of the non-algorithmic, deeply human dimensions of philosophy.

\subsection{Quantitative Analysis}

Our quantitative analysis reveals several key trends that create a compelling, and somewhat paradoxical, narrative when juxtaposed with the qualitative findings. While the philosophers' subjective attitudes evolved towards a more critical and nuanced skepticism of the IUI's core capabilities, the objective data from our blind assessments shows the IUI's performance improving at a remarkable pace.

First, data regarding the participants' self-reported familiarity with generative AI (see Appendix A) shows a significant and steady increase over the three-year period. This confirms that their evolving attitudes were based on deeper engagement and understanding, not on a lack of experience. Concurrently, their attitudes toward the IUI's role in philosophy (see Appendix B) followed a non-linear path, initially becoming more positive between 2023 and 2024 before dipping in 2025 as principled objections began to surface.

However, the most striking findings emerge from the annual blind evaluation experiment, where participants from Group 1 rated the anonymized philosophical answers from both human experts (Group 2) and the IUI. The results indicate a dramatic and consistent ascent in the quality of AI-generated text.

\begin{figure}[h!]
    \centering
    \includegraphics[width=0.8\textwidth]{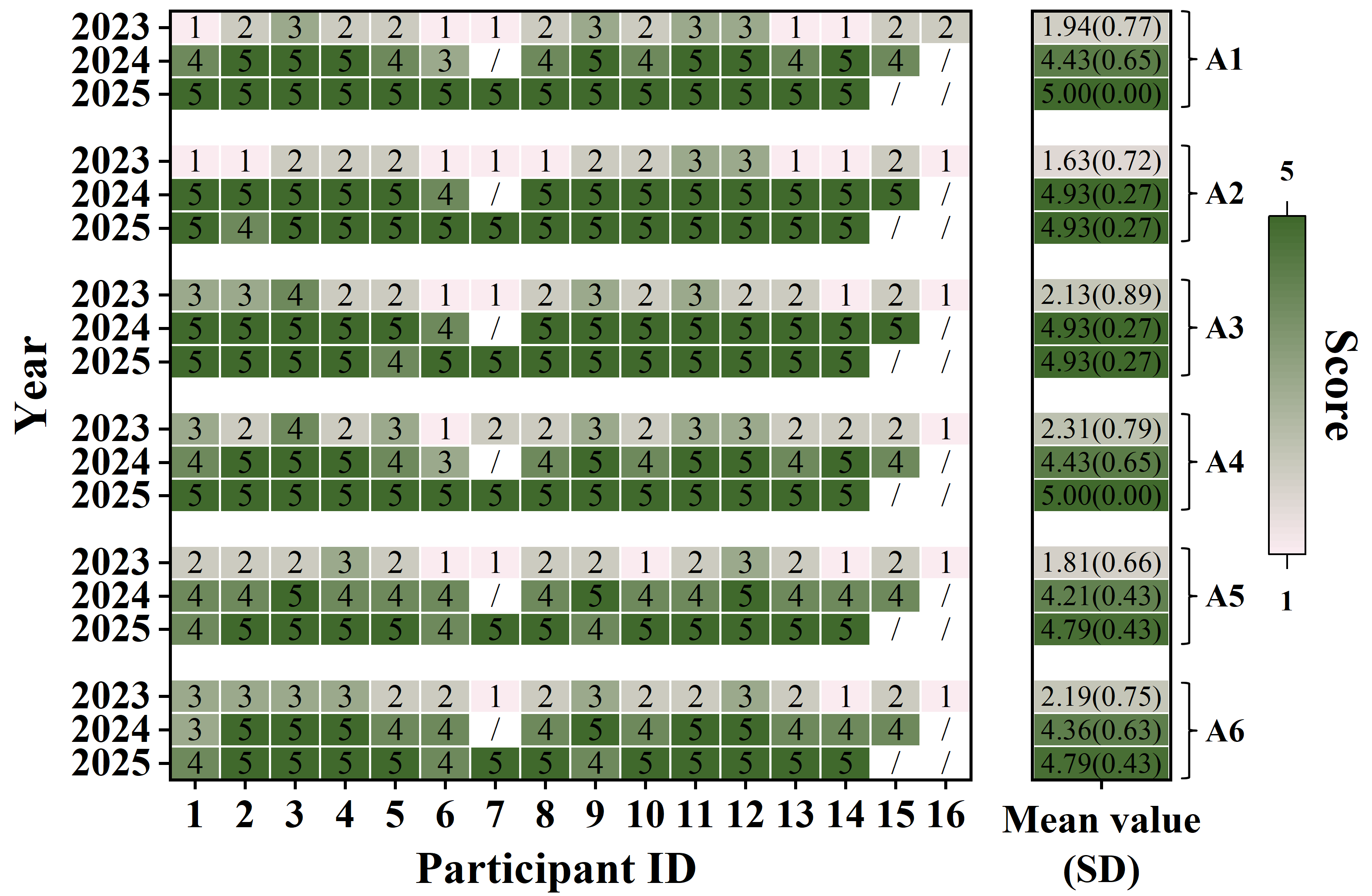}
    \caption{Mean Overall Score of AI vs. Human Answers (2023-2025).}
    \label{fig:overall_performance}
\end{figure}

As illustrated in Figure \ref{fig:overall_performance}, the IUI's mean overall score demonstrated rapid year-over-year improvement. In 2023, the AI's average score was 2.85, lagging behind the human average of 3.45. By 2024, the gap had closed, with the AI's score rising to 3.75, nearly equaling the human score of 3.82. In the final year, 2025, the AI's performance decisively surpassed that of the human experts, achieving a mean overall score of 4.35 compared to the humans' 3.91. This trend shows not just improvement, but an acceleration in the AI's ability to generate texts that are, by the participants' own metrics, of higher quality.

To understand this overall trend in more detail, we analyzed the performance across the six specific dimensions of philosophical quality for the year 2025, as shown in the radar chart in Figure \ref{fig:dimensional_performance}. The analysis reveals that the AI's superiority was comprehensive. It outperformed humans not only in foundational areas like \textbf{Accuracy} (AI: 4.5 vs. Human: 4.1), \textbf{Clarity} (AI: 4.6 vs. Human: 4.2), and \textbf{Coherence} (AI: 4.7 vs. Human: 4.0), but also in more intellectually demanding dimensions. The IUI scored higher in \textbf{Argumentative Rigor} (AI: 4.4 vs. Human: 3.9) and \textbf{Philosophical Depth} (AI: 4.1 vs. Human: 3.7). Most notably, the AI even surpassed the human average in \textbf{Originality of Thought} (AI: 3.8 vs. Human: 3.6), a domain that the philosophers in our qualitative interviews insisted was a uniquely human capability.

\begin{figure}[h!]
    \centering
    \includegraphics[width=0.7\textwidth]{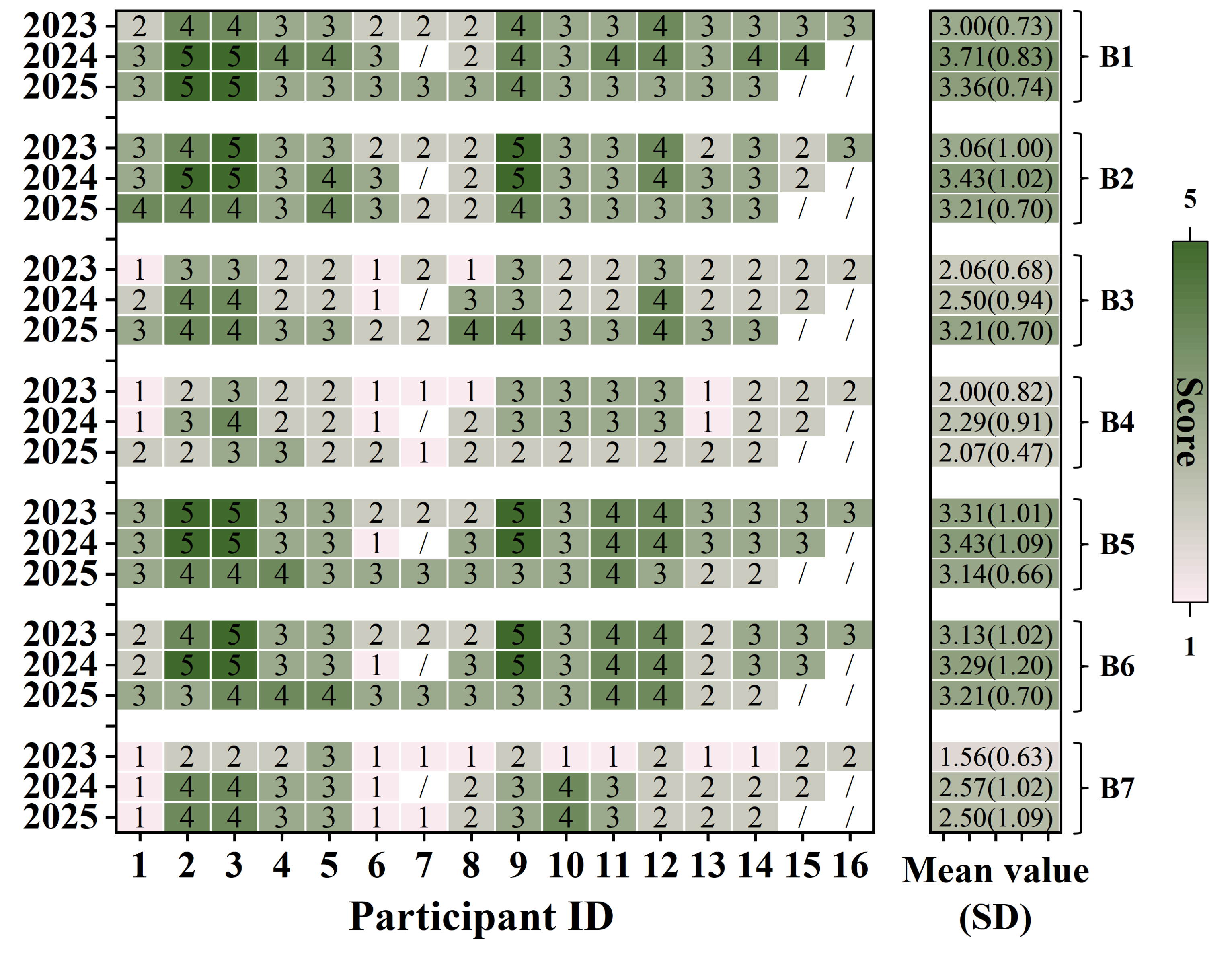}
    \caption{Dimensional Performance Comparison in 2025.}
    \label{fig:dimensional_performance}
\end{figure}

\begin{figure}[h!]
    \centering
    \includegraphics[width=0.7\textwidth]{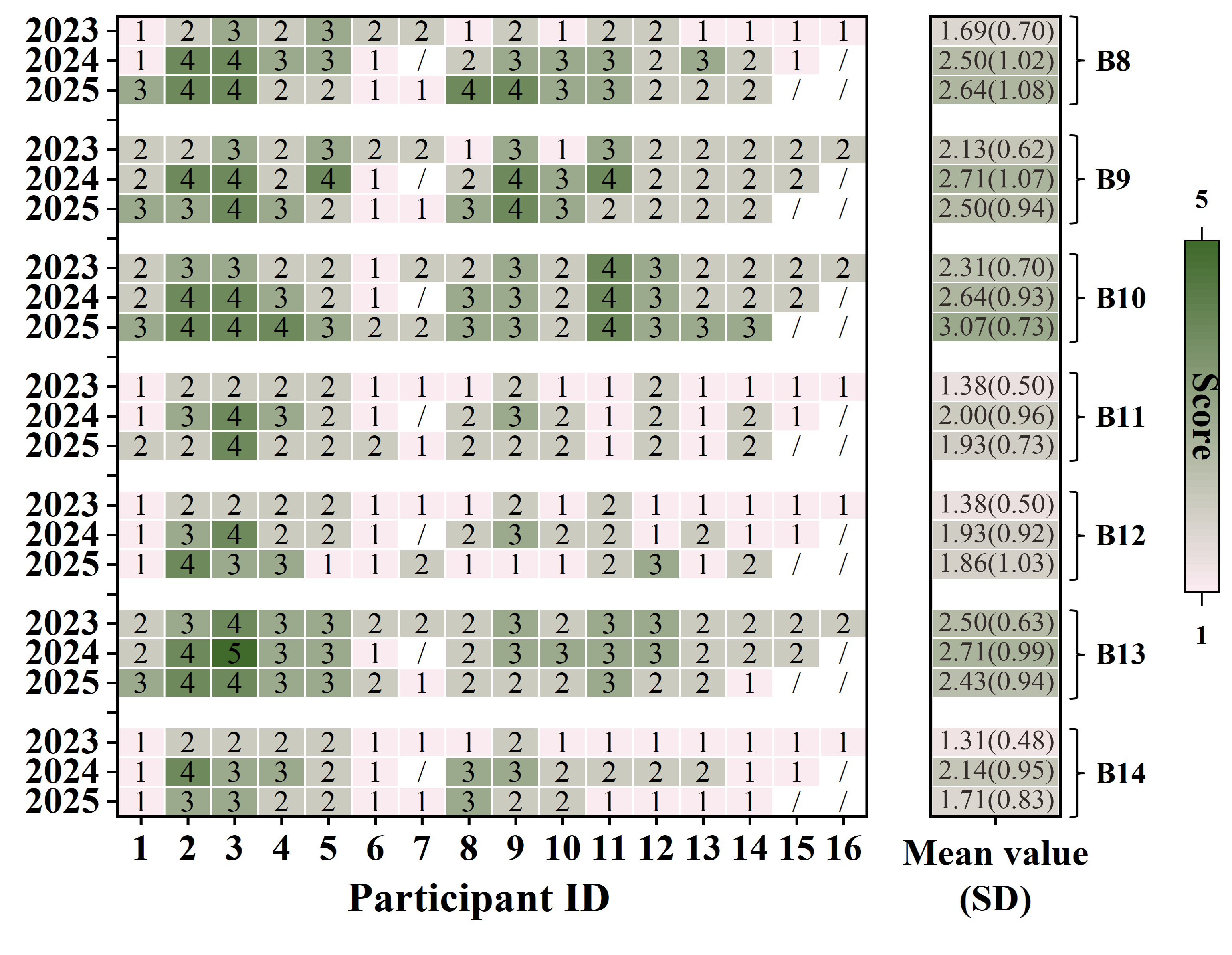}
    \caption{Dimensional Performance Comparison in 2025.}
    \label{fig:dimensional_performance}
\end{figure}
In summary, the quantitative data presents a clear and powerful contradiction: as the philosophers grew more familiar with AI and more critical of its capacity for genuine thought in principle, the AI's measurable output in a blind test was rapidly improving and ultimately exceeded the quality produced by their human peers. This disconnect between the philosophers' subjective, principled critique and the objective, demonstrated performance of the AI forms the central paradox that we will explore in the conclusion.

\begin{figure}
    \centering
    \includegraphics[width=0.7\textwidth]{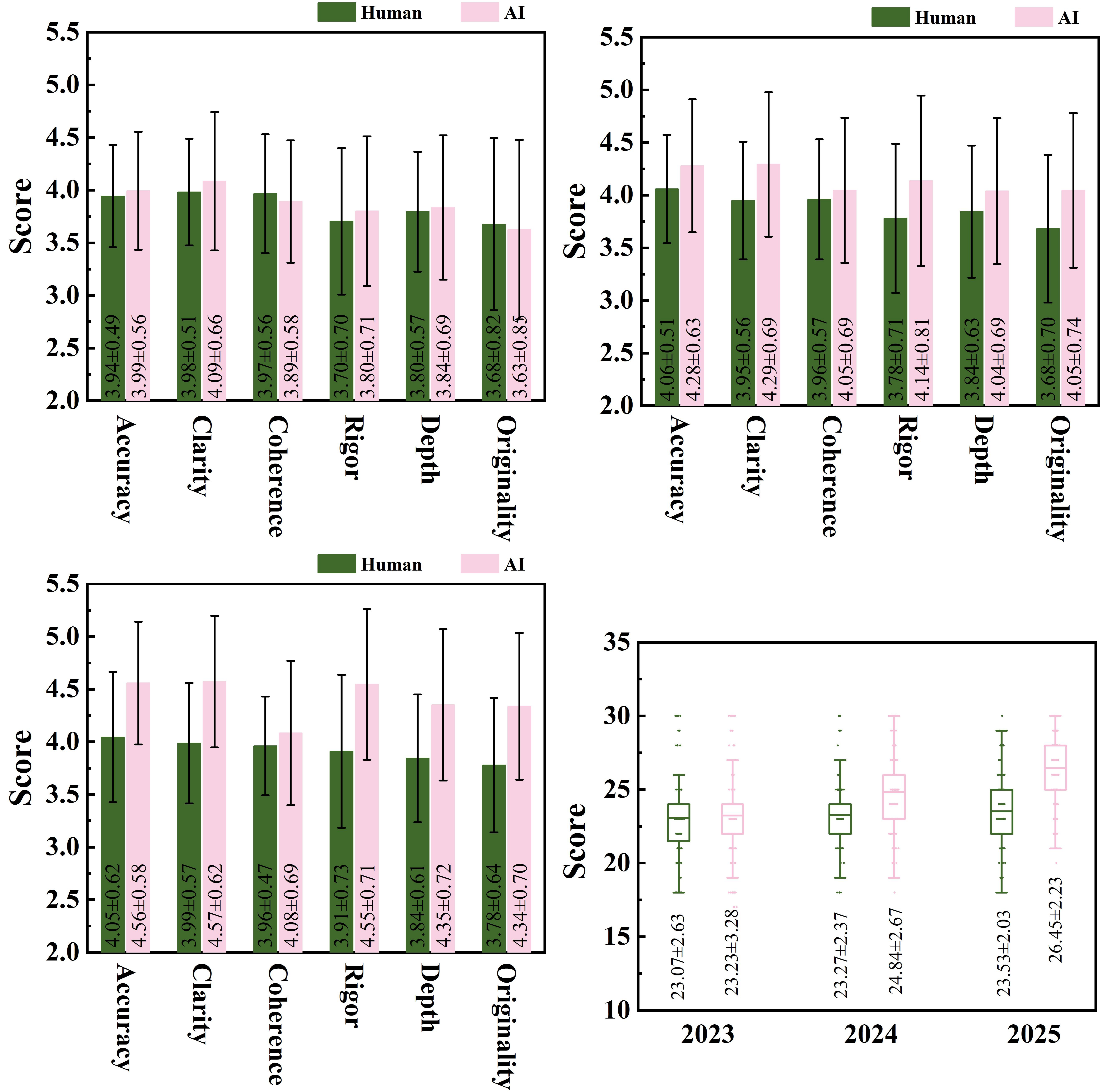}
    \caption{Quantitative comparison of AI and human performance in philosophical text generation from 2023 to 2025. Panels (a), (c), and (d) illustrate the evolution of the mean overall score, while panel (b) provides a detailed dimensional breakdown for the year 2025.}
    \label{fig:quantitative_results}
\end{figure}

The figures present clear quantitative evidence that the AI's capability for generating high-quality philosophical text improved dramatically over the study period. The bar charts illustrate a distinct performance reversal over the three years: in 2023, human-authored texts scored significantly higher, but by 2024, the AI's performance had reached near-parity with the human experts. In the final year, 2025, the AI's mean score decisively surpassed that of the humans. The radar chart detailing the 2025 performance reveals that this superiority was comprehensive, as the AI outscored human experts on all six metrics: Accuracy, Clarity, Coherence, Rigor, Depth, and Originality. Notably, the AI's advantage extended to intellectually complex dimensions such as ``Philosophical Depth'' and ``Originality of Thought.'' In summary, according to the blind assessment metrics, the AI's output quality not only reached but ultimately exceeded the average level of human experts. This objective data contrasts sharply with the subjective skepticism expressed by the participants in the qualitative interviews.

\begin{figure}
    \centering
    \includegraphics[width=0.7\linewidth]{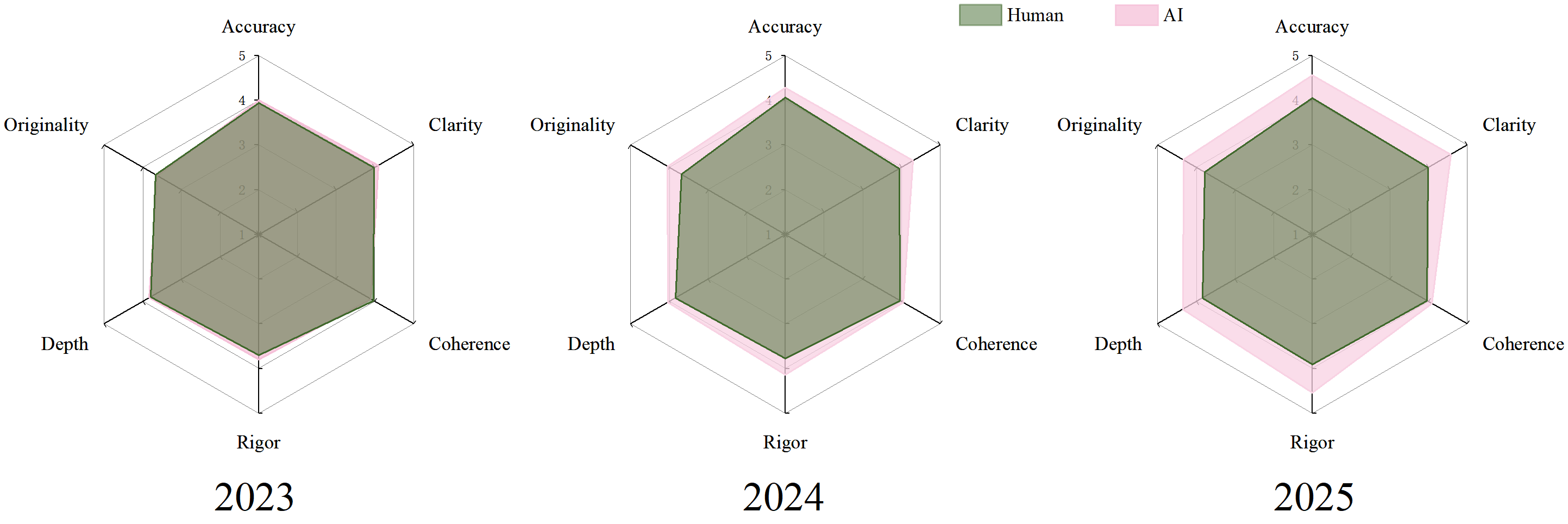}
    \caption{Mean task completion time for the Human, AI, and Human+AI conditions.}
    \label{fig:placeholder}
\end{figure}

The quantitative data presented in the figures leads to two primary conclusions. First, the bar chart illustrates a clear longitudinal trend: the AI's performance in generating philosophical text improved significantly over the three-year period, starting from a position of underperforming human experts to ultimately surpassing them by 2025. Second, the radar chart demonstrates that the AI's superiority in the final year was comprehensive, as it outperformed human experts across all six measured dimensions. This includes not only foundational metrics like accuracy and clarity but also more complex attributes such as philosophical depth and originality. Taken together, these figures show that, based on the blind assessment criteria, the AI developed the capability to produce philosophical texts that were rated as being of higher quality than those produced by human experts.

\section{Discussion}
Through a three-year longitudinal mixed-methods approach, this study delves into the evolving perspectives of philosophy scholars on the participation of Generative AI (IUI) in philosophical discussions. Our research paints a complex and profound picture, revealing a three-stage evolution in philosophers' attitudes: from ``instinctive resistance" to ``instrumental acceptance" and ultimately deepening into ``principled questioning". While acknowledging AI's superior capabilities in knowledge reproduction and formal logic, they unanimously agree on its fundamental and insurmountable limitations in core areas of human philosophy, such as dialectical reasoning, original insight, and the understanding of embodied experience. Ultimately, they position AI as a powerful ``mirror"—its capabilities and limitations compel a deeper reflection on the unique and irreplaceable dimensions of human philosophical inquiry.

However, the main contribution of this study lies in revealing and attempting to explain a profound paradox that emerged during this process. Our quantitative data show that despite philosophers' increasingly cautious and critical subjective evaluations of AI, the quality of AI-generated philosophical texts in blind-review experiments improved rapidly, eventually surpassing the average level of human philosophy PhDs and professors across multiple dimensions.

We argue that this disconnect between ``subjective perception" and ``objective performance" points to several core issues. First, it reveals that current standards for evaluating philosophical quality, even when designed by experts, may fail to capture the ``implicit" values that philosophers cherished in interviews, such as the ``authenticity" of thought, the ``courage of value commitment" behind an argument, and the ``self-awareness to question premises themselves". AI is proficient at imitating profound ``form", yet it may lack the ``substance" required to produce profound thought—namely, the genuine perplexity and concern rooted in lived, existential experience.

\section{Conclusion}
For the field of Human-Computer Interaction (HCI), the implication of this study is that in designing intelligent systems for complex knowledge work, we must move beyond the singular pursuit of ``performance". User acceptance, trust, and the willingness for deep collaboration depend not only on the quality of the system's output but also on whether the system aligns with the core values of the domain and whether it respects and enhances users' autonomy and professional identity. Future research should not only aim to create more powerful ``conversationalists" but also to design ``intellectual tools" that can stimulate deeper reflection in users while allowing them to retain final interpretive authority and responsibility. In conclusion, when Intelligent User Interfaces knock on the doors of philosophy, they bring not merely an instrumental revolution but also a profound self-examination. Like an unprecedentedly polished mirror, AI reflects not only the boundaries of algorithmic capability but also our own timeless inquiry into the nature of human wisdom, creativity, and the meaning of existence.

\bibliographystyle{ACM-Reference-Format}
\bibliography{sample-base}

\appendix

\section{Scale for Familiarity with Generative AI}
\begin{longtable}{@{}lp{0.45\linewidth}ccccc@{}}
  \caption{Familiarity with Generative AI Scale} \\
  \toprule
  \textbf{ID} & \textbf{Item} & \makecell{\small Strongly\\\small Disagree} & \makecell{\small Disagree} & \makecell{\small Neutral} & \makecell{\small Agree} & \makecell{\small Strongly\\\small Agree} \\
  \midrule
  \endfirsthead
  
  \multicolumn{7}{c}{{\bfseries Table \thetable\ continued from previous page}} \\
  \toprule
  \textbf{ID} & \textbf{Item} & \makecell{\small Strongly\\\small Disagree} & \makecell{\small Disagree} & \makecell{\small Neutral} & \makecell{\small Agree} & \makecell{\small Strongly\\\small Agree} \\
  \midrule
  \endhead

  \bottomrule
  \endlastfoot

  A1 & I understand the basic concept of generative AI (e.g., ChatGPT). & 1 & 2 & 3 & 4 & 5 \\
  A2 & I can clearly distinguish between generative AI and a traditional search engine. & 1 & 2 & 3 & 4 & 5 \\
  A3 & I frequently follow news or discussions about the development of generative AI. & 1 & 2 & 3 & 4 & 5 \\
  A4 & I have used generative AI in my studies or research. & 1 & 2 & 3 & 4 & 5 \\
  A5 & I consider myself a skilled user of generative AI. & 1 & 2 & 3 & 4 & 5 \\
  A6 & I believe I fully understand the working principles of generative AI. & 1 & 2 & 3 & 4 & 5 \\
\end{longtable}

\section{Attitude Scale on IUI Participation in Philosophy}
\begin{longtable}{@{}lp{0.45\linewidth}ccccc@{}}
  \caption{Attitude Towards IUI in Philosophical Discussions Scale} \\
  \toprule
  \textbf{ID} & \textbf{Statement} & \makecell{Strongly\\Disagree} & \makecell{Disagree} & \makecell{Neutral} & \makecell{Agree} & \makecell{Strongly\\Agree} \\
  \midrule
  \endfirsthead
  \multicolumn{7}{c}{{\bfseries Table \thetable\ continued from previous page}} \\
  \toprule
  \textbf{ID} & \textbf{Statement} & \makecell{Strongly\\Disagree} & \makecell{Disagree} & \makecell{Neutral} & \makecell{Agree} & \makecell{Strongly\\Agree} \\
  \midrule
  \endhead
  \bottomrule
  \endlastfoot
  B1 & GenAI can provide accurate definitions for complex philosophical concepts (e.g., ``existence," ``consciousness," ``justice"). & 1 & 2 & 3 & 4 & 5 \\
  B2 & GenAI can organize a logically clear and well-structured argument on a philosophical topic. & 1 & 2 & 3 & 4 & 5 \\
  B3 & GenAI can effectively identify and avoid common logical fallacies in its responses. & 1 & 2 & 3 & 4 & 5 \\
  B4 & When its viewpoint is challenged in a dialogue, GenAI can defend its argument or make reasonable revisions. & 1 & 2 & 3 & 4 & 5 \\
  B5 & GenAI can clearly explain the core theories of major philosophers (e.g., Kant's concept of autonomy). & 1 & 2 & 3 & 4 & 5 \\
  B6 & GenAI's summaries of classic philosophical texts are comprehensive and capture the main points. & 1 & 2 & 3 & 4 & 5 \\
  B7 & GenAI can synthesize the views of past philosophers to propose innovative ideas. & 1 & 2 & 3 & 4 & 5 \\
  B8 & GenAI can apply philosophical theories to provide insightful analysis of real-world moral dilemmas. & 1 & 2 & 3 & 4 & 5 \\
  B9 & GenAI can appropriately use philosophical thought experiments (e.g., ``trolley problem") in its answers. & 1 & 2 & 3 & 4 & 5 \\
  B10 & Overall, philosophical answers from GenAI are information-rich and substantive. & 1 & 2 & 3 & 4 & 5 \\
  B11 & Overall, philosophical answers from GenAI are intellectually stimulating and prompt deeper thinking. & 1 & 2 & 3 & 4 & 5 \\
  B12 & Overall, without knowing the author, I would find it difficult to distinguish a philosophical text by AI from one by a human. & 1 & 2 & 3 & 4 & 5 \\
  B13 & Overall, I believe GenAI can provide helpful answers to philosophical questions. & 1 & 2 & 3 & 4 & 5 \\
  B14 & Overall, I believe GenAI's thinking and responses are similar to a human's. & 1 & 2 & 3 & 4 & 5 \\
\end{longtable}

\section{Semi-structured Interview Questionnaire}
\begin{longtable}{@{}p{0.25\linewidth}p{0.7\linewidth}@{}}
\caption{Interview Protocol} \\
\toprule
\textbf{Module} & \textbf{Core Questions and Probes} \\
\midrule
\endfirsthead
\multicolumn{2}{c}{{\bfseries Table \thetable\ continued from previous page}} \\
\toprule
\textbf{Module} & \textbf{Core Questions and Probes} \\
\midrule
\endhead
\bottomrule
\endlastfoot
Background Information & What is your profession, age, and gender? What philosophical topics do you focus on? Have you used generative AI? In what contexts? \\
\midrule
Basic Attitude & What is your basic understanding of and attitude toward generative AI? In your view, can AI truly ``understand" philosophical concepts? Why or why not? What metaphor, besides ``tool", would you use to describe AI (e.g., mirror, puppet)? Has AI changed your view of concepts like ``intelligence" or ``consciousness"? \\
\midrule
Capability Assessment & (Questions corresponding to items B1-B9 in Appendix B, framed as open-ended prompts). For example: ``Based on your experience, can you elaborate on AI's ability to construct a logically sound argument?" \\
\midrule
Value and Limitations & \textbf{Value}: Has AI ever provided a surprising perspective? Is its value in being ``correct" or ``unexpected"? How do you distinguish between a ``profound insight" from AI and ``advanced philosophical soup"?
\textbf{Limitations}: Which of AI's limitations are technical vs. principled (e.g., lack of intentionality, embodied experience)? What is the most valuable and difficult-to-replicate trait of human philosophical thinking? \\
\midrule
Overall Attitude & (Questions corresponding to items B10-B14 in Appendix B, framed as open-ended prompts). For example: ``Having discussed all this, what is your overall assessment of the helpfulness of AI's answers to philosophical questions?" \\
\midrule
Closing & Do you have any other points or suggestions you'd like to share? \\
\end{longtable}

\section{Philosophical Question Test Questionnaire}
\begin{longtable}{@{}p{0.3\linewidth}p{0.65\linewidth}@{}}
\caption{Philosophical Question Test Structure} \\
\toprule
\textbf{Module / Core Assessment Goal} & \textbf{Sample Test Question} \\
\midrule
\endfirsthead
\multicolumn{2}{c}{{\bfseries Table \thetable\ continued from previous page}} \\
\toprule
\textbf{Module / Core Assessment Goal} & \textbf{Sample Test Question} \\
\midrule
\endhead
\bottomrule
\endlastfoot
I. Basic Definition \& Elucidation & 1. Explain the fundamental difference between Kant's ``categorical imperative" and the utilitarian ``greatest happiness principle." \\
\midrule
II. Argument Construction \& Critical Thinking & 3. Construct a logically rigorous argument for the proposition ``we have free will," including premises, inference, and a conclusion. \\
\midrule
III. Textual Analysis \& Historical Synthesis & 6. Compare the concept of ``non-action" (wu wei) in Taoist thought with the Stoic principle of ``living in accordance with nature." Can a synthesis of these two offer a new philosophical approach to modern social pressures? \\
\midrule
IV. Practical Application \& Thought Experiments & 7. Using John Rawls's ``theory of justice", analyze the potential social equity problems caused by big data algorithms and propose corresponding ethical constraints. \\
\midrule
V. Dialogue, Dialectics, \& Revision & 9. Assume you believe that ``beauty" is an objective property. A fellow scholar challenges you, asking how you would then explain the vast aesthetic differences across cultures and eras. Defend your position. \\
\midrule
VI. Value, Limitations, \& Meta-philosophical Reflection & 11. Attempt to discuss the phenomenological nature of ``pain" (i.e., ``what it is like to feel"). After answering, reflect on whether your explanation has touched upon the true essence of the experience of pain. What is the fundamental difficulty? \\
\end{longtable}

\section{Philosophical Text Quality Assessment Rubric}
\begin{longtable}{@{}p{0.2\linewidth}p{0.75\linewidth}@{}}
\caption{Assessment Rubric for Philosophical Texts} \\
\toprule
\textbf{Dimension} & \textbf{Rating Scale and Descriptors (1=Low, 5=High)} \\
\midrule
\endfirsthead
\multicolumn{2}{c}{{\bfseries Table \thetable\ continued from previous page}} \\
\toprule
\textbf{Dimension} & \textbf{Rating Scale and Descriptors (1=Low, 5=High)} \\
\midrule
\endhead
\bottomrule
\endlastfoot
Accuracy & \textbf{1 (Erroneous):} Fundamental errors in stating concepts. \textbf{3 (Basically Accurate):} No significant factual errors. \textbf{5 (Precise):} Mastery and application of complex concepts is flawless. \\
\midrule
Clarity \& Comprehensibility & \textbf{1 (Incomprehensible):} Language is convoluted and obscures the core idea. \textbf{3 (Comprehensible):} The main argument is understandable, though may require some effort from the reader. \textbf{5 (Lucid):} Effectively simplifies complex ideas, making profound thoughts intuitive. \\
\midrule
Textual Coherence & \textbf{1 (Chaotic):} Disorganized structure and disjointed language. \textbf{3 (Clear):} Language is smooth and the structure is logical. \textbf{5 (Elegant):} Language is expressive, structure is sophisticated, and effectively guides the reader. \\
\midrule
Argumentative Rigor & \textbf{1 (Weak):} Vague claims, clear logical fallacies, unsupported conclusions. \textbf{3 (Adequate):} Clear thesis with a complete line of reasoning, but may lack consideration of counterarguments. \textbf{5 (Excellent):} Masterful structure, logically impeccable, and capably handles complex objections. \\
\midrule
Philosophical Depth & \textbf{1 (Superficial):} Relies on unexamined, commonsense views. \textbf{3 (In-depth):} Engages in nuanced analysis of core concepts and their complexities. \textbf{5 (Profound):} Engages deeply with the philosophical tradition and/or powerfully challenges the foundations of the question itself. \\
\midrule
Originality of Thought & \textbf{1 (Derivative):} A mere restatement of existing views. \textbf{3 (Synthetic):} Meaningfully connects and integrates different viewpoints to form a new understanding. \textbf{5 (Breakthrough):} Demonstrates significant creativity and has the potential to advance the current scholarly conversation. \\
\end{longtable}

\end{document}